\documentstyle[epsf,fleqn,aps]{revtex}
\begin{document}
\tighten                        
\twocolumn[\hsize\textwidth\columnwidth\hsize\csname 
@twocolumnfalse\endcsname

\title{Coarsening dynamics of adsorption processes 
with diffusional relaxation}
\author{M.  D. Grynberg}

\address{Departamento de F\'{\i}sica, Universidad Nacional 
de La Plata, C.C. 67, (1900) La Plata, Argentina}
\maketitle


\begin{abstract}

We investigate the late coarsening stages of 
one dimensional adsorption processes with diffusional relaxation.
The nonequilibrium domain size distribution is studied by means of
the field theory associated to the stochastic evolution.
An exact asymptotic solution satisfying dynamical scaling is given
for cluster sizes smaller than the average domain length.
Our results are supported and compared with Monte Carlo simulations.

\vspace{10 pt}
\hspace{0.5cm} Published in Phys. Rev. {\bf E} 57, 74 (1998).
\vspace{-12 pt}
\end{abstract}
\vskip2pc] \narrowtext

Random sequential adsorption models have been systematically
investigated as basic prototypes of monolayer growth 
in many physical, chemical and biological processes
\cite{Evans,Privman}. 
The characteristic feature dominating the late stage 
dynamics of such 
phenomena is the jamming of the available area of deposition, 
leading to the formation of partially covered 
and fully blocked states.
Recent experimental advances indicate that even for large colloidal 
particles, monolayer deposits  may further redistribute on the substrate 
by particle diffusion on time scales comparable with the adsorption process
\cite{Ramsden}.
A range of theoretical efforts including exact solutions, 
asymptotic methods and
extensive numerical simulations based on simple microscopic models,
has been used to understand the role of fluctuations and 
collective effects in these processes \cite{Evans,Privman}.
There is ample simulational evidence for the existence of a scaling 
regime where the system is effectively made up of pure phase regions
separated by narrow interfaces, highly reminiscent 
of quenched binary alloys
and fluids at low temperatures \cite{Gunton}. For large times $t$ 
a network of domains emerges such that can be characterized by a 
{\it single} length scale ${\cal L} (t)$ namely, 
the average domain size, 
which coarsens continuously. On general grounds, 
typical statistical quantities 
are expected to be scaling functions 
of a single argument involving both 
space and time \cite{Bray}.

Although there are several exact results in coarsening dynamics
available mostly in one dimension \cite{Bray}, 
they essentially refer to the dynamical scaling 
of  two-point correlations (structure  factor),
and average properties of the  domain size distribution (DSD). 
More complete descriptions of cluster 
growth at the submicron level clearly
require knowledge of the DSD itself, 
a quantity of fundamental interest 
in modern nucleation theories and accessible 
to light microscope studies \cite{Gunton}.
As a contribution in this direction,
here we present an asymptotic analysis of nonequilibrium DSD
in a simpler system lending itself more readily 
for this calculation and still
capturing basic aspects of coarsening phenomena. 

Specifically, we consider an extension of the 
random dimer deposition 
problem of Flory \cite{Evans} where 
{\it vacant} pairs of nearest-neighbor lattice sites 
are filled randomly by two hard core particles 
at a time, say with adsorption rate 
$R$. To prevent an otherwise jamming behavior (resulting from both 
hard core interactions and the lack of 
nearest-neighbor vacancies), we enable
the system to relax diffusively by single particle hopping between
nearest neighbors with probability $h$, though yet 
avoiding multiple occupancy.
This leads to an effective hopping motion
of vacant sites which recombine to form larger voids 
accessible to deposition
attempts, ultimately covering the full crystalline 
limit at large times.

Turning to  the evaluation of DSD, in what follows we 
shall restrict our discussion 
to one-dimensional systems. Far from being trivial, 
asymptotic solutions
in $d = 1$ share many features emphasized in higher dimensions
and do provide a demanding test for theories of late 
stage coarsening,
particularly for the dynamical scaling hypothesis.
The strategy is to study the field theory
that can be associated to the master equation of 
our adsorption-diffusion process \cite{Yo}.
This can be achieved by means of a (pseudo) fermionic 
representation in which its occupation numbers 1 or 0 at site $j$ 
correspond to particle or vacancy at that location.
After introducing creation (annihilation) fermi operators 
$C^{\dag}_j$ ($C_j$) along with the local density fields
$\hat{n}_j \equiv   C^{\dag}_j C_j$, the
stochastic evolution at a given time can be represented 
by the action 
$e^{- H t}$ of the quantum 'Hamiltonian' 
{\mathindent = 0cm \begin{eqnarray}
\nonumber
 H  = &-&R \sum_j C^{\dag}_j C^{\dag}_{j+1} 
 -h \sum_j \left( C^{\dag}_j C_{j+1} +{\rm h.c.}\right) + N  R\\
\label{Hamilton}
 &+& (R - 2h) \sum_j \hat{n}_j \hat{n}_{j+1}
 +2 (h - R) \sum_j \hat{n}_j\,,
\end{eqnarray} 
operating over a periodic chain with $N$ locations.}
Here, deposition (hopping) of dimers (particles) 
is described by the effect of
the first (second) sum in Eq.\,(\ref{Hamilton}), 
whereas conservation of 
probability requires the action of the remaining 
(diagonal) field operators.
We address the reader to Ref. \cite{Yo} for a more 
detailed derivation in this and related systems.

The analysis of DSD requires to consider the number
of domains having  at least $L$ consecutive 
particles (with $L$ arbitrarily large), 
by averaging over all possible histories up to a certain instant.
For a given initial probability 
distribution $\vert P(0) \rangle$,  this is related to the 
$L$-point correlation function \cite{Kawasaki}
\begin{equation}
\label{correlator}
F (L,t)  = \frac{1}{N}\,\sum_j \langle \psi^{^{\!\!\!\!\sim}} 
\vert \,\hat{n}_{j+1}\,...\hat{n}_{j+L}\,e^{- H t}\,
\vert P(0)\rangle\,,
\end{equation}
where $\langle \psi^{^{\!\!\!\!\sim}} \vert$ is an equally 
weighted sum of 
all accessible configurations, i.e. the {\it left} 
steady state of $H$. 
Although  the diagonalization of the evolution operator
becomes fairly standard  by choosing $R = 2 h$\,, i.e.
dimer adsorption and particle diffusion occuring with the
same probability \cite{comment}, the difficulties associated to 
the evaluation of these high order correlators
as we shall see,  are simplified significantly 
by {\it detaching} dimers with rate $\epsilon \equiv 2h - R$, 
whether or not the selected pair of adjacent  particles 
arrived together.
Though this fictitious process introduces additional terms 
in Eq.\,(\ref{Hamilton}) \cite{Yo}, the latter constraint 
ensures that $H$ remains bilinear in $C,\, C^{\dag}$ operators.
Therefore, it can be readily verified that 
after a Bogoliubov-type similarity 
transformation in momentum space \cite{Lieb},  
we are finally left with the 
free fermion Hamiltonian
\begin{equation}
\label{diag}
H = \sum_{- \pi < q < \pi} \lambda_q \xi^+_q 
\xi_q\:,\:\:\lambda_q = b + a \cos q\,,
\end{equation}
where $a = R - \epsilon,\, b = R + \epsilon$, and the elementary
$\xi$-excitations are given by
{\mathindent = 0cm \begin{eqnarray}
\nonumber
\vec \xi_q &=& - \frac{e^{i \frac{\pi}{4}}}{\sqrt{N}}\,\sum_j 
e^{i q j}  \pmatrix{ - \alpha\,\cos\,\theta_q\, & \!\!\!
 i \,\alpha^{-1}\, \sin\,\theta_q \cr \alpha\,
 \sin\,\theta_q\, & \!\!\! 
 i \,\alpha^{-1}\,\cos\,\theta_q } \vec C_j\,,\\
\label{bogo}
 \alpha &=& (R/\epsilon)^{1/4}, \;\; 
\tan \theta_q  = \alpha^2  \cot \frac{q}{2}\,,
\end{eqnarray}
with $\vec \xi_q$, $\vec C^{\dag}_j$ denoting respectively
$\xi^+_q \choose \xi_{-q}$, $C^{\dag}_j \choose C_j$.}
So, in the limit $\epsilon \to 0$ in which the 
original process is recovered,
the dynamical evolution becomes critical as it 
is dominated asymptotically 
by low-lying massless modes $q_0 = \mp \pi \pm q$ with spectrum 
$\lambda_{q_0} \propto q^2$.

We are especially interested to elucidate the long time behavior of
the correlators (\ref{correlator}) for which it is 
convenient to expand the initial probability
distribution in terms of these fermions. In particular,  
starting from an empty substrate, 
it is a simple matter to check that $\vert P(0) \rangle$ 
corresponds to the coherent pair state
\begin{equation}
\label{BCS}
\vert P(0) \rangle = \prod_{0 < q < \pi}
\left(\,1\,+\,\tan \theta_q\: \xi^+_q\,\xi^+_{-q}\,
\right)\,\vert\,\psi\,\rangle\,,
\end{equation}
where $\vert \psi \rangle$ is the right 
vacuum (steady) state of $H$. 
Hence, from Eqs.\,(\ref{correlator}) and 
(\ref{diag}) it follows that
for nonvanishing desorption rates, $F(L,t)$ can be expanded
perturbatively in powers of 
$u \equiv e^{- 4 \epsilon \, t}$ as 
$\rho_s^L + \frac{1}{N}\sum_j 
\sum_{n =1}^L  F_{n , j} (L,t)$,  where
{ \mathindent = 0cm \begin{eqnarray}
\nonumber
F_{n ,j} (L,t) &=& \frac{u^n}{n!} \sum_{q_1} ...  \sum_{q_n}\, 
\langle \psi^{^{\!\!\!\!\sim}} \vert 
C^{\dag}_{j+1} C_{j+1} \, ... \,C^{\dag}_{j+L} C_{j+L}\\
\label{typical}
&\times&  \prod_{i=1}^n\, e^{- 2 a  (1+ \cos q_i)\, t} \, 
\tan \theta_{q_i}\,
 \,\xi^+_{q_i} \: \xi^+_{-q_i} \,
\vert  \psi  \rangle\,,
\end{eqnarray} 
$0 < q_i < \pi$ and,  $\rho_s = 1/(1+\sqrt{\epsilon/R})$ is  
the coverage of the steady state.}
To evaluate the vacuum expectation value 
of this product, we use Wick's theorem \cite{Negele} 
for which we compute all 
pair contractions (in this case, 
steady state expectation values) contributing
to such typical term. The six kind of contractions 
that occur are readily obtained
if we combine the inverse of  Eq.\,(\ref{bogo}) along
with its associated anticommutative algebra. In the limit  
$N\to \infty$ this finally yields
\begin{eqnarray}
\label{contraction1}
 \langle C_l \, C_m \rangle & = & - \langle C^{\dag}_l \, 
C^{\dag}_m \rangle =
\frac{\sqrt{ R \, \epsilon}}{2  b}\,(1+\beta^2) \, \beta^{n-1}\,,\\
\label{contraction2}
\langle C^{\dag}_l C_m \rangle &=& - \langle C_l  \, 
C^{\dag}_m \rangle = - \langle C_l \, C_m \rangle\,,\\
\label{contraction3}
\langle C_l \: \xi^+_q \rangle &=& 
\frac{e^{i \frac{\pi}{4}}}{\sqrt{N}}\,
\cos \theta_q \, e^{i q l}\,,\\
\label{contraction4}
\langle C^{\dag}_l \,\xi^+_q \rangle &=& 
\frac{e^{-i \frac{\pi}{4}}}{\sqrt{N}}\,\sin \theta_q\,
e^{i q l}\,,
\end{eqnarray}
where $n = m - l > 0$, and  
$\beta = (\sqrt{\epsilon} - \sqrt{R})/(\sqrt{\epsilon} + \sqrt{R})$.
Since $\vert P(0) \rangle$ has zero total momentum, notice that
the sum over all these pairings in Eq.\,(\ref{typical}) 
(with their corresponding permutation signature), 
results independent of the site location,  i.\,e. $F_{n , j} (L,t) 
\equiv F_n (L,t)$. Thus, $F(L,t)$ remains translationally
invariant for all subsequent times, as it should. 

Clearly, for {\it finite} detaching rates 
$\epsilon$, there is an exponentially large number of 
pairing groups contributing 
to Eq.\,(\ref{typical}). Even the calculation of the leading 
order $F_1$ results prohibitively involved. 
However, in the limit $\epsilon \to 0$  contractions 
(\ref{contraction1}), (\ref{contraction2})
and (\ref{contraction3}) vanish as $\sqrt{\epsilon}$.
Thus, by taking into account  the Bogoliubov angles 
appearing in Eqs.\,(\ref{bogo}) 
and (\ref{typical}), a moment of reflection shows that there 
are only two relevant pairing forms 
contributing to $F_1$ namely, $\langle C^{\dag} \xi^+ 
\rangle^2\, \langle C C \rangle\,$,
e.g. $\langle C^{\dag}_l \xi^+_q \rangle \langle 
C^{\dag}_m \xi^+_{-q} \rangle
\langle C_l C_m \rangle$,  and $\langle C^{\dag} \xi^+ 
\rangle \, \langle C \xi^+ \rangle\,$,
a remarkable simplification.
Using the  {\it multiplicity} and signature of these products, 
and after introducing the  integrals
\begin{equation}
\label{integrals}
f^{\pm}_n (\tau) = \frac{1}{\pi}\,\int\limits_0^{\pi}\,
\frac{e^{-\,\tau\,\cos q}}{\sin q}\,
\sin n q\: (1\,\pm\,\cos q)\,dq\,,\\
\end{equation}
it is straightforward to show that
{ \mathindent = 0cm \begin{equation}
\label{order1}
F_1 (L,t) = - e^{- \tau}\!  \left[ L I_0 (\tau) + \!
\sum_{n=1}^{L-1} (-1)^n (L -n) f^+_n (\tau) \right],
\end{equation} 
where $I_0 (\tau)$ is a modified Bessel function 
of the first kind \cite{Gradshteyn}, and $ \tau \equiv 2 R t$.}
Similarly, the number of  pairing groups which yield 
a net contribution to higher 
orders of Eq.\,(\ref{typical}) remains bounded 
irrespective of the domain size $L$, 
though proliferating very rapidly with the order $n$. 
For instance, there are 24 
products of the form $\langle C \xi^+ \rangle^2 \, 
\langle C^{\dag} \xi^+ \rangle^2$, 
72 $\langle C^{\dag} \xi^+ \rangle^4 \, \langle C C \rangle^2$, and
72  $\langle C^{\dag} \xi^+ \rangle^3 \, \langle C \xi^+ \rangle \,
\langle C C \rangle$ to be considered in the calculation of 
$F_2 (L, t)$. The analysis is simple albeit in fact rather lengthy.
In the long time limit and {\it fixed} domain size, 
it can be shown that the former 24 products contribute 
as $L^2/t$, whereas the latter 144
are bounded by $L^5/t^3$.  More specifically, employing the
integrals (\ref{integrals}) along with modified 
Bessel functions of integer order $I_n$, we find
{ \mathindent = 0cm \begin{eqnarray}
\nonumber
F_2 (L,t) \!\! &\sim& 
\frac{e^{-2 \tau}}{2}\,\left\{ \, L\, (L-1) \,I^2_0 (\tau) - 
\sum_{n=1}^{L-1} \, (L -n)  \right. \\
\label{order2}
 &\times&   \left.  
\phantom{\!\!\!\!\!\!\!\!\!\!\!\!\!\sum_{n=1}^{L-1} }
 \left[ I_n^2 (\tau) + f^+_n (\tau)\,f^-_n (\tau)\, 
 \right]\, \right\} +
{\cal O} (L^5 / t^3)\,.
\end{eqnarray} }

We are not concerned here with the possibility of improving the
second order calculation, which is a problem of great technical
difficulty, but in showing that the approach, even in lowest order,
can be successfully applied to late coarsening stages.
It should be borne in mind however, that for {\it arbitrarily large} 
domain sizes the prefactors involved in higher orders
of Eq.\,(\ref{typical}) result  increasingly weighted. Nevertheless,
it turns out that these contributions become irrelevant within
the scaling regime  $t \to \infty$, 
$L \to \infty$, with $L^2/t \ll 1$, 
where they provide solely subdominant large-time corrections. 

In studying this asymptotic region  it is helpful to consider 
the  number $N_L (t)$ of  filled 
$L$-intervals between two vacancies, 
along with the density of domains $N_d(t)$ 
averaged up to a given instant.
Clearly, the probability  to observe 
a cluster having {\it exactly} $L$ 
particles at that time is $P(L,t) = N_L (t) /N_d(t)$. 
It can be easily checked that   $N_L = F(L) + F(L+2) - 2 F(L+1)$, 
$\forall t$, while on the other hand 
$N_d$ coincides with the number 
of particle--vacancy interfaces, and therefore
can be calculated as $\langle \,
\hat{n}_j\,(1\!-\!\hat{n}_{j+1}) \rangle$, 
(the brackets indicate an average over histories).
Following a similar analysis discussed as in \cite{Yo},
$N_d (t)$ can be shown  to yield  $e^{-\tau} I_1 (\tau)$. 
Hence, by virtue of the asymptotic behavior of 
Eqs.\,(\ref{order1}) and (\ref{order2}),
it finally turns out that for $z \equiv L  
/\sqrt{2 \pi \tau} \ll 1$, 
$P(L,t)$ satisfies the dynamical scaling hypothesis, namely 
$P (L,t) = {\cal P}(z) /\sqrt{2 \pi \tau}$,  where
${\cal P} (z)$ is a universal scaling  function given by
\begin{eqnarray}
\nonumber
{\cal P} (z) &=& \frac{\pi}{2}\, z \, 
e^{-\,\pi\, z^2}\,\left[\,1 +
 {\rm erfc} (\sqrt {\pi}\, z)\,\right]\\
\label{scaling1}
&+&\, \frac{1}{2}\, \left( 1- e^{-2 \pi z^2} \right) + 
{\cal O} (z^5)\,,
\end{eqnarray} 
and ${\rm erfc}\,(x)  = \frac{2}{\pi} \int_x^{\infty} 
\! e^{-u^2}\! du$ is
the complementary error function \cite{Gradshteyn}. 
Thus, there is an emerging typical length scale 
${\cal L}(t) = \sqrt {2 \pi \tau}$ 
which characterizes the whole domain structure at large times. 
In a statistical sense, the domain morphology
becomes self similar if all lengths are measured 
in units of  ${\cal L}(t)$. In fact,
this characteristic scale can be ultimately identified with 
the average domain size,
since by construction $\langle L \rangle  = \rho(t)/N_d(t)$, where 
$\rho (t) =  \sum_L L N_L (t)  \to 1$,  is the particle coverage; 
so  in the long time limit $\langle L \rangle  \equiv {\cal L}(t)$. 
Moreover, it is known \cite{Yo}
that {\it two} point vacancy-vacancy 
correlation functions $C(L,t) \! = \! \frac{1}{N}\! \sum_j \langle (1\!-\!\hat{n}_j)
(1\!-\!\hat{n}_{j+L}) \rangle$  scale asymptotically as
{ \mathindent = 0cm \begin{equation}
\label{scaling2}
C(L,t) = \frac{e^{-\,\pi\, z^2}}{ {\cal L}^2(t)}\, 
\left[\,\frac{\pi}{2}\, z \,
{\rm erfc} (\sqrt {\pi}\, z) \,+\,2 \sinh ( \pi \, z^2)\right] \,,
\end{equation}  
where the scaling parameter is taken as in Eq\,(\ref{scaling1}).}
Thus, we see that both average 
domain size and pair correlation length
coalesce into a {\it single} physical scale which is typically 
diffusional. This is in line with the coarse grained 
(hydrodynamic) level of description, the so called
(noiseless) model A or time dependent Ginzburg-Landau 
approach \cite{Gunton,Langer}, 
in which there is a single non-conserved scalar field 
(in our case, the particle density)
leading to a characteristic scale which grows as $\sqrt {t}$.
In addition, these results reveal a close asymptotic 
relationship between DSD and two point correlations, namely
\begin{equation}
P(L,t) = 2 {\cal L} (t) C(L,t) + {\cal O } (L^2/t^{3/2})\,,
\end{equation}
so, apart from a global change of scale they closely 
follow each other.

We have conducted Monte Carlo simulations to confirm the validity
of our theoretical expectations in a 
periodic chain of $N = 10^5$ sites. The microscopic dynamical rules 
accounting for the stochastic process 
described by Eq.\,(\ref{Hamilton}), 
are as follows. Starting from an empty lattice, dimer deposition 
attempts on randomly targeted 
{\it bonds} are made with probability $R$ while maintaining 
single occupancy throughout.
Alternatively, a particle hopping attempt with probability $h$
takes place isotropically within the selected bond 
provided it contains a vacant site;
otherwise the move is rejected. The unit Monte Carlo 
step is defined such that 
each bond is checked once on average. 
This corresponds to $N$ trials per unit time.
We direct the reader's attention to Fig. 1, 
where we display the DSD results obtained
for a wide range of domain sizes, after averaging 
over $3 \times 10^4$  histories
up to $10^3$ steps. This has been adequate to suppress numerical 
fluctuations arising particularly from large sizes $L$ 
yet smaller than the typical system length.

As expected, by setting $R = 2h$ our results 
for $L \ll {\cal L} (t)$  
reproduce completely the asymptotic scaling distribution
(\ref{scaling1}). However, for arbitrarily large 
domain sizes the relevance of the 
high order corrections referred to above, 
evidently is reflected  in the progressive departure between
theory and simulation. Nevertheless, our approach turns out to be 
still successful to yield an accurate estimate of the 
most probable cluster size
$\sim 0.468 \, {\cal L}(t)$,  occurring in fact within a regime 
of intermediate lengths.
For  $L \gg {\cal L} (t)$ we content ourselves with giving 
just the numerical results
displayed in the inset of Fig.\,1 which suggest  the 
DSD follows an exponential distribution 
scaling as ${\cal P} (z) \sim  P_0 \, e^{ - k z}$, 
with $k = 1.3(1)$ and $P_0 = 1.8(1)$.
Similar results were observed starting 
with other initial conditions, the scaling
distribution always appearing in the long time limit, 
as long as only short-range 
spatial correlations are initially present. 
It is worth remarking that 
this robustness applies as well to 
the unrestricted general case $R \ne 2 h$
(also shown in Fig. 1), where the dynamics 
can not be solved explicitly.

The existence of dynamic scaling, however, 
appears to be associated with 
a clean separation between fast microscopic 
time scales $\propto 1/R$
and slow collective modes, such as 
the gapless $\xi$-excitations of Eq.\,(\ref{bogo}).
For instance, no scaling behavior seems to hold  
for finite detaching rates $\epsilon$.
In fact,  the non-critical dynamics 
includes a subcase ($\epsilon = R = h$) 
entirely soluble by standard transfer 
matrix techniques \cite{Yo}, in which there
is no dynamic scaling of any kind.
Thus, the issue of universality in critical 
dynamics arises immediately. 
Whether or not slightly different 
nonequilibrium systems share a 
similar set of exponents and scaling 
functions is still an open problem
which is receiving systematic attention \cite{Droz}. 
In this context, we conclude
by examining a number of common aspects 
between the dynamics discussed so far
and an alternative process of cluster growth 
on a lattice, in which hard-core 
particles diffuse and eventually give birth to another 
particle at an adjacent site \cite{Isaac}. 
At the level of the average 
domain size, it is by now well established that both processes 
coarsen diffusively \cite{Droz,Isaac}.
Furthermore, even the dynamics of two point correlations
can be described asymptotically by the {\it same} scaling function 
(\ref{scaling2}) \cite{Yo,Isaac}.
However, at the more demanding microscopic level of DSD 
universality no longer holds. In fact \cite{Isaac},  
the birth process follows a scaling distribution 
${\cal P} (z) = \frac{\pi}{2} z e^{-\frac{\pi}{4} z^2}$,
whose gaussian tail indicate the occurrence 
of relatively larger domains
(see Fig. 1), whereas on the other hand it 
can not  be either rescaled into
Eq.\,(\ref{scaling1}) beyond third order in $L/{\cal L} (t)$. 

In summary, we have presented a 
scaling picture which accounts for
the late coarsening stages of simple adsorption processes
where, however,  fluctuation-induced behavior is essential.
As often in nonequilibrium statistical mechanics, even
the solution of the simplest models helps 
to convey a clearer understanding
of the many characteristics present in complex systems.  
While progress has been accomplished in $d = 1$,  a similar 
understanding of spatial structures in higher dimensional 
systems still requires further investigations.

It is a pleasure to thank R. B. Stinchcombe for helpful
discussions and remarks. The author acknowledges
financial support of CONICET, Argentina.


\vspace{-3.5cm}
\begin{figure}
\hbox{%
\epsfxsize=3.9in
\epsffile{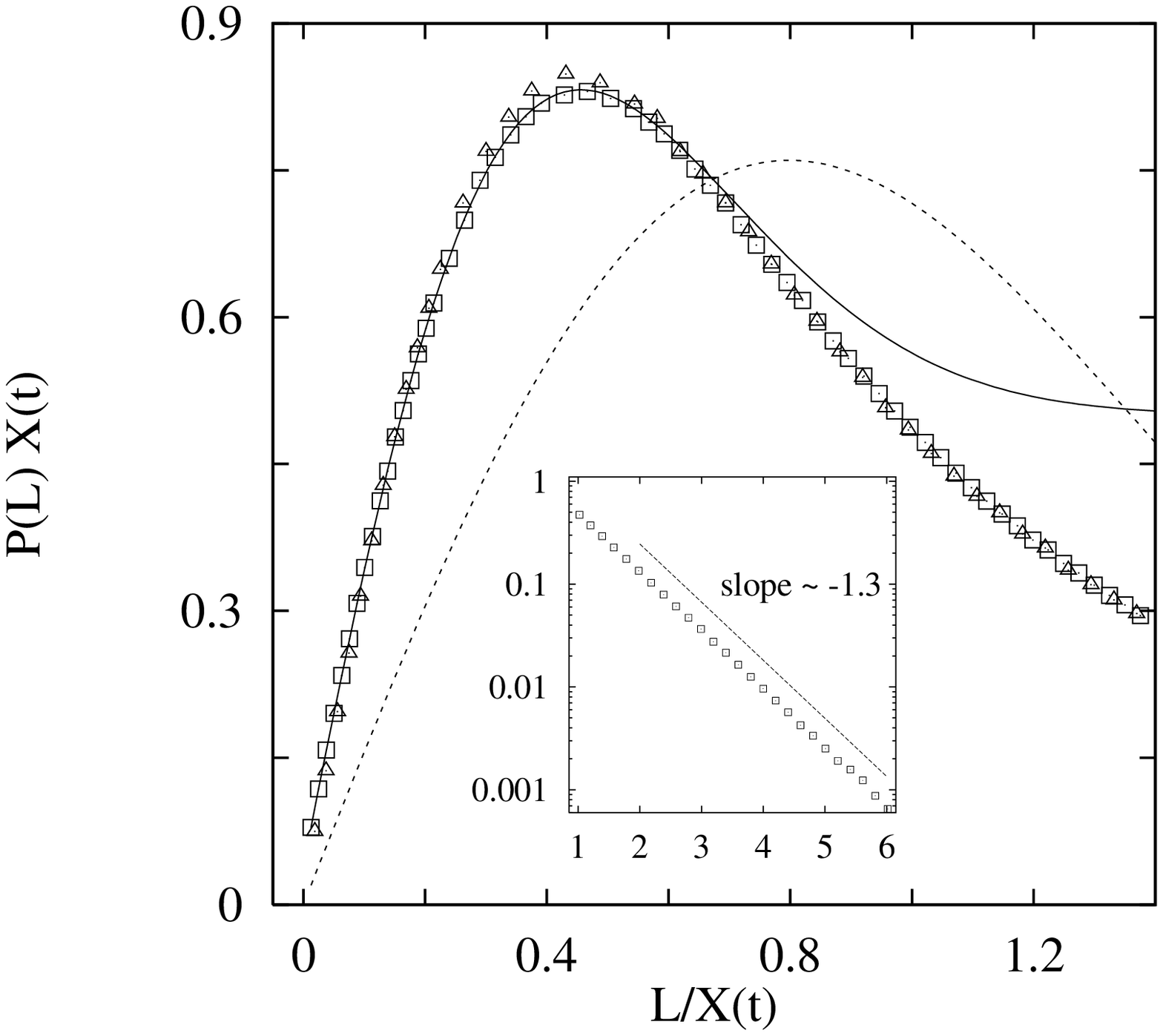}}
\vspace{-2.5cm}
\caption{Asymptotic domain size 
distribution at $t = 10^3$ for $R = 1, h= 0.5$
(squares), and $R = 1, h = 0.1$ (triangles, 
non-soluble case).  The averages
were taken over $3 \times 10^4$ histories 
starting from an empty chain 
of $10^5$ sites. For domain sizes $L$ smaller 
than the average domain size
$\propto t^{1\!/2}$ [here denoted as $X(t)$], 
the numerical data follow closely
the theoretical results given by Eq.(14) in the 
text (solid line). For comparison,
we show the scaling distribution of the birth process 
discussed in the text
(dashed line). The inset provides evidence of 
exponential distribution for large domain sizes.}
\end{figure}


\begin{references}

\bibitem{Evans} For comprehensive review and literature list, 
consult J. W. Evans, Rev. Mod. Phys. {\bf 65}, 1281 (1993).

\bibitem{Privman}V. Privman, {\it Annual Reviews 
of Computational Physics},
edited by D. Stauffer (World Scientific, 1995), Vol. 3 
and references therein.

\bibitem{Ramsden} J.J. Ramsden, J. Stat. Phys. 
{\bf 73}, 853 (1993);
C. J. Murphy {\it et al.}, Science, {\bf 262}, 1025 (1993).

\bibitem{Gunton} J. D. Gunton, M. San Miguel, P. S. Sahni, 
{\it Phase 
Transitions and Critical Phenomena}, edited by C. Domb and 
J. L. Lebowitz (Academic Press, London, 1983), Vol. 8

\bibitem{Bray} The current status of the field has 
been reviewed by
A. J. Bray, Adv. Phys. {\bf 43}, 357 (1994).

\bibitem{Yo} M.D. Grynberg  and R.B. Stinchcombe, 
Phys. Rev. E {\bf 52},  
6013 (1995); M. D. Grynberg, 
{\it Annual Reviews of Computational Physics},
edited by D. Stauffer (World Scientific, 1996), Vol. 4;
F. Alcaraz {\it et al.}, Ann. Phys. (NY) {\bf 230}, 135 (1994);
G. M. Sch\"utz, J. Phys. A {\bf 28}, 3405 (1995), Phys. Rev.
E {\bf 53}, 1475 (1996); A. A. Lushnikov, Sov. Phys. JETP
{\bf 64}, 811 (1986).


\bibitem{Kawasaki} K. Kawasaki, 
{\it Phase Transitions and Critical Phenomena}
edited by C. Domb and M.S. Green 
(Academic Press, London 1972), Vol. 2.
See also Ref. \cite{Yo}.

\bibitem{comment} Although this 
relationship might appear rather  limitative, 
more general Monte Carlo simulations 
show no substantial change in 
the asymptotic dynamics, suggesting
that the case $R = 2h$ is actually generic.

\bibitem{Lieb}  Consult for instance, 
E. H. Lieb, T. D. Schultz, and D. C. 
Mattis, Ann.  Phys. (NY), {\bf 16},  407 (1961).

\bibitem{Negele}See for example J. W. Negele and H. Orland, 
{\it Quantum Many-Particle Systems}, (Addison-Wesley,1987).

\bibitem{Gradshteyn} I.S. Gradshteyn and I.M. Ryzhik, 
{\it Table of Integrals, Series and Products}, Fifth Edition, 
edited by A. Jeffrey  (Academic Press, 1994).

\bibitem{Langer} J. S. Langer, {\it Solids far from Equilibrium}, 
edited by C. Godr\`eche, (Cambridge University Press, 1992); 
P. C. Hohenberg and B. I. Halperin, 
Rev. Mod. Phys. {\bf 49}, 435 (1977).

\bibitem{Droz}  M. Droz and L. Sasvari, 
Phys. Rev. E {\bf 48}, R2343
(1993); D. Balboni, P.A. Rey and M. Droz, 
Phys. Rev. E {\bf 52}, 6220
(1995); M. Henkel, E. Orlandini, and G.M. Sch\"utz, 
J. Phys. A {\bf 28}, 6335 (1995).

\bibitem{Isaac} For a recent account on this
and related processes see D. ben-Avraham, 
{\it Nonequilibrium Statistical Mechanics 
in One Dimension}, edited by V. Privman, 
(Cambridge University Press, 1996), 
and references therein.

\end{references}
\end{document}